\def\ref{\par\noindent\hang}

\def\etal{{et al\ }}

\def\spose#1{\hbox to 0pt{#1\hss}}
\def\approxlt{\mathrel{\spose{\lower 3pt\hbox{$\sim$}}
	\raise 2.0pt\hbox{$$<$$}}}
\def\approxgt{\mathrel{\spose{\lower 3pt\hbox{$\sim$}}
	\raise 2.0pt\hbox{$>$}}}

\def\multleft#1{\hbox to size{\vbox {\halign {\lft{##}\cr #1}}\hfill}\par}
\def\multright#1{\hbox to size{\vbox {\halign {\rt{##}\cr #1}}\hfill}\par}

\def\$<${\thinspace}
\def\s{\hbox{\phantom{5}}}	

\def\boxit#1{\vbox{\hrule\hbox{\vrule\kern3pt\vbox{\kern3pt
          #1 \kern3pt}\kern3pt\vrule}\hrule}}

\def\arcm{{\arcm\thinspace arcmin}}

\def\km{{\rm\thinspace km}}

\def\Mpc{{\rm\thinspace Mpc}}

\def\s{{\rm\thinspace s}}

\def\kmps{\hbox{$\km\s^{-1}\,$}}

\def\kmpspMpc{\hbox{$\kmps\Mpc^{-1}$}}

\documentstyle[psfig]{mn}
\begin{document}
\hsize=6truein

\title[]{On the mass distribution in the Shapley Supercluster inferred 
from X-ray observations} 

\author[]
{\parbox[]{6.in} {S. Ettori, A.C. Fabian and D.A. White\\
\footnotesize
Institute of Astronomy, Madingley Road, Cambridge CB3 0HA \\
}}                                            
\date{accepted, March 1997}
\maketitle

\begin{abstract}
We present an analysis of a mosaic of {\it ROSAT} PSPC and {\it
Einstein Observatory} IPC
X--ray observations of 14 clusters and 2 groups of galaxies, enclosed in a
sky area $15^\circ \times20^\circ$ centred on A3558 within the Shapley
Supercluster region.

From the mass of each cluster, extrapolated
to a density contrast of 500, we define 4 large structures:
(1) the core of the Shapley Supercluster with radius 13 $h_{50}^{-1}$ Mpc; 
(2) the core plus A1736; (3) the core, A1736 and 
the western extension (A3528-A3530-A3532)
and A3571; and (4), adding the northern cluster pair A1631--A1644 
to structure (3), 
the Supercluster as a whole enclosed within a radius $\sim 90
h_{50}^{-1}$ Mpc. The observed total masses range between 
3.5--8.5 $\times 10^{15} M_{\odot}$. The mass values derived from 
the observed intracluster gas (assuming a
baryon fraction consistent with primordial nucleosynthesis) are of the
order of few times $10^{16} M_{\odot}$. Given these
estimates, the core is a bound structure with a very
significant overdensity of
at least 5 times the critical density, indicating that is 
approaching maximum expansion before collapsing. Structure (2)
has an overdensity of 1.7 on a scale of $\sim 30 h^{-1}_{50}$ Mpc.
The core is then a 2--3.7 $\sigma$ fluctuation in an initial Gaussian
perturbation field, normalized for galaxies and clusters. 
The highest value applies if we extrapolate to a density 
contrast of 200 and assume that $\Omega_{\rm b} < 0.095$.

The baryon fraction of the core of the Shapley Supercluster,
with A1736, is about 15 per cent over a radius of 28 $h^{-1}_{50}$ Mpc.

\end{abstract}


\section{INTRODUCTION} 

First noted by Shapley (1930), the Shapley Supercluster is a large 
overdensity of galaxies in the region of
Centaurus--Hydra ($\alpha_{2000}: 13^h25^m, \delta_{2000}: -30^o$,
$z \sim 0.046$) which has been widely studied both in the 
optical (Melnick \& Moles 1987; Vettolani \etal 1990; 
Raychaudhury \etal 1991; Quintana \etal 1995) and X-ray wavebands (Day \etal
1991 based on GINGA data; Breen \etal 1994 on {\it Einstein
Observatory} IPC data).

Analysis of the structure of the supercluster,
performed by Zucca \etal (1993; SC 26 in their catalogue) and 
Einasto \etal (1994; SC 80), suggest that it is a concentration
with more than 20 clusters of galaxies 
within a volume of $60\times100\times200$ Mpc$^3$ 
($\sim 50$ times the average density of ACO clusters at the same $|b|
\sim 30^o$, Vettolani \etal 1990; we assume $H_0 = 50$ \kmpspMpc 
and $q_0 = 0.5$), distributed
in 4 different regions: (i) the core around A3558 and A3562; (ii) an
eastern part centred around
A3571; (iii) a western region with A3532 and
(iv) an elongation to the north (A1736, A1644). The unusual nature of this
region is also borne out through X--ray observations which show these
six clusters to be amongst the 46 X--ray brightest at high galactic
latitude ($|b| > 20^\circ$, Edge \etal 1990). Furthermore,
the Shapley concentration probably contributes about 10--20 per cent to
the optical dipole observed in the motion of the Local Group with
respect to the Cosmic Microwave Background 
(Lynden-Bell \etal 1988, Raychaudhury 1989, Scaramella \etal 1989).

The Shapley Supercluster is thus an extremly dense region where peculiar 
and global cluster--supercluster  characteristics can be studied. 
Following an initial analysis by Fabian (1991), which showed that 
its central region is the largest mass overdensity yet discovered 
on so a large scale, we attempt here to estimate the matter 
(gas, stellar, gravitational) distributions, the total baryon 
fraction, and whether the region is bound.
To this aim, we collate all the available {\it ROSAT} PSPC images 
of the clusters  present of the Shapley area, completing the resulting
mosaic with {\it Einstein Observatory} IPC images of 3 other clusters.
Then, their surface brightness profiles are deprojected to estimate the
gas and gravitational masses,  as described in Section 2. Combining 
these results with constraints on the mass in galaxies, 
we obtain limits on the baryon fraction of the Shapley
Supercluster in Section 3. In Section 4 we discuss the matter 
(gas and dark) distribution in the various 
Shapley structures, and the implications of our results for cosmology and 
large scale structure. We summarize our conclusions in Section 5.

\section{X-RAY DATA AND ANALYSIS} 

The surface brightness profiles from
{\it ROSAT} Position Sensitive Proportional Counter
(PSPC) images of 11 clusters and 2 groups, and
{\it Einstein Observatory} Imaging Proportional Counter (IPC) 
data for 3
other clusters (A1631, A1644, A1736), are analysed using the image
deprojection procedure, as described below. 

The spatial position of the PSPC and IPC fields 
and the clusters of the ACO catalogue suspected as members of the Shapley 
Supercluster are shown in Fig.~1; 
a mosaic of the exposure--corrected images is shown in 
Fig.~2. 

\begin{figure}
\psfig{figure=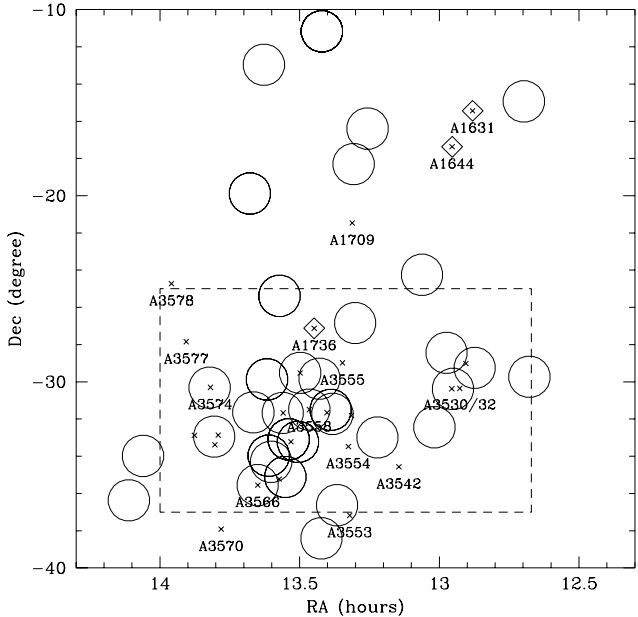,width=0.5\textwidth,angle=0}
\caption{The spatial distribution of the 47 PSPC circular fields, each
with a diameter $2^\circ$, of the 3 IPC 75$'$-square field 
and the positions of the ACO clusters are shown enclosed within a radius
of $15^{\circ}$ from $13^{\rm h}28^{\rm m}$ and $-25^{\circ}15'$.
The dashed line indicates the region shown in Fig.~2.}\end{figure}

In Table~1 the ACO coordinates and 
details of the X-ray exposure for each cluster in the sample
are summarized. The optical data on redshift $z$ and velocity dispersion
$\sigma$, are taken from the literature; the X--ray data includes the
instrument, the exposure time, and the spatial average intracluster 
medium (ICM) temperature, $T_{\rm X}$, (available from previous work). 
In the last column, we quote the Galactic absorption taken from 
the 21-cm determinations by Stark \etal (1992).
 
\begin{figure*} \raisebox{20.ex}{
\parbox[b]{0.3\textwidth}{\caption{In the panel below, we present a mosaic
of the X--ray
observations. In the panel to the right, we show the observations
of clusters in the northern extension (A1631 and A1644) which is located
$\sim 16^{\circ}$ away from
the core in the NW direction. Note the two
groups SC1327-312 and SC1329-313 
(west and east, respectively) between A3558 and A3562. A close--up image
of the core is shown in Fig.~7.} } } \
\psfig{figure=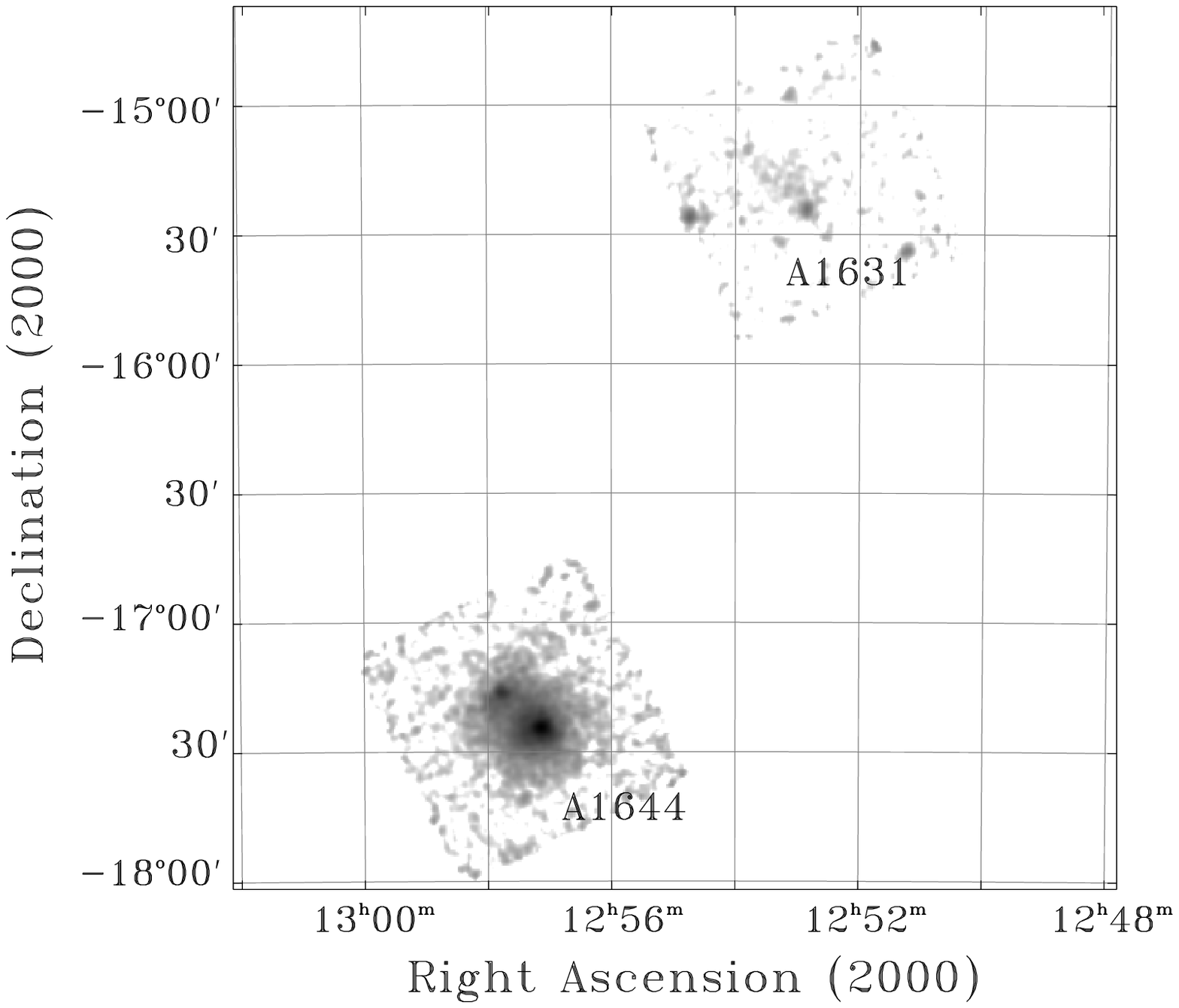,width=0.5\textwidth,angle=0}
\psfig{figure=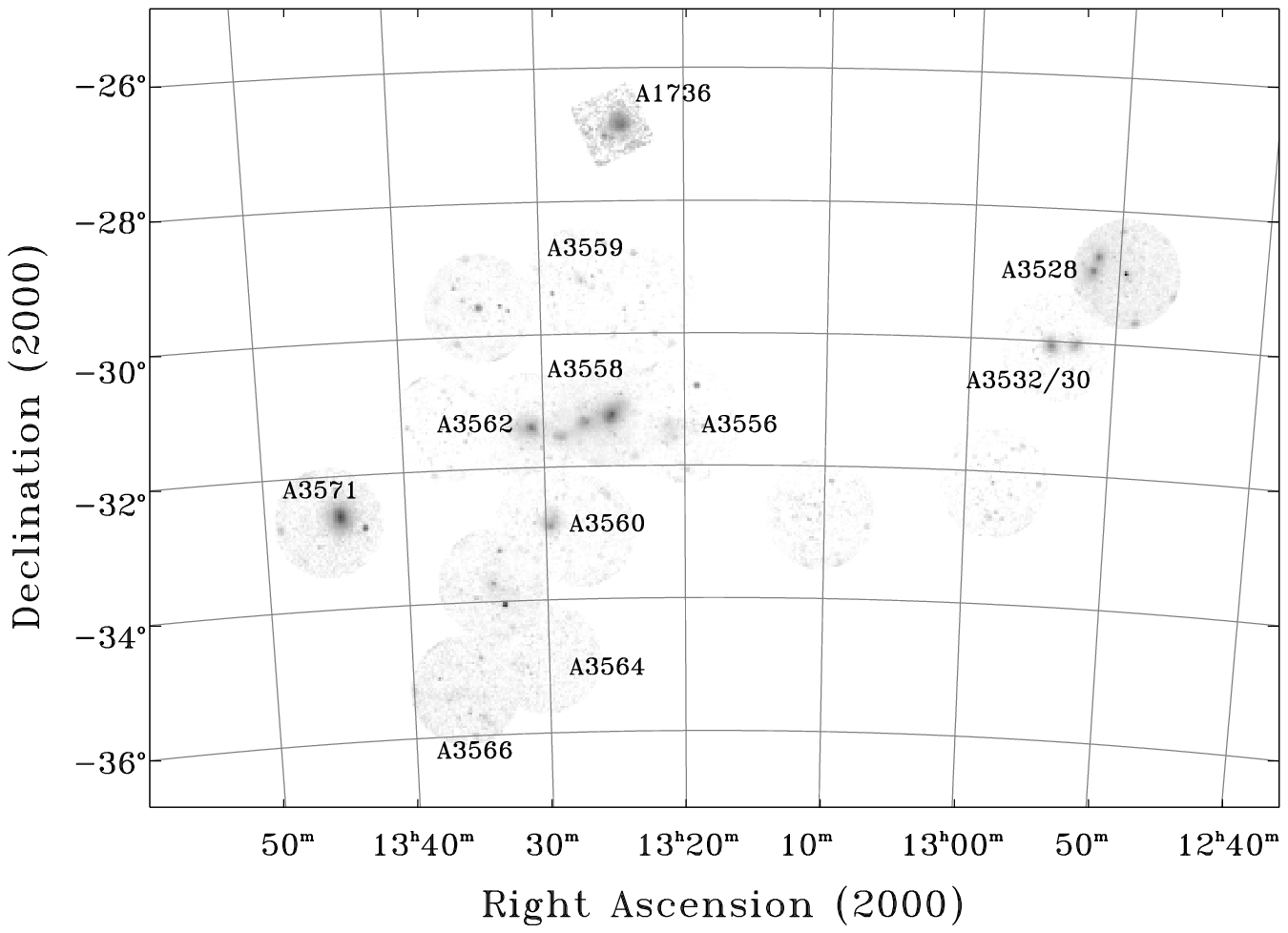,width=\textwidth,angle=0}
\end{figure*}

\begin{table*}   \caption{The X-ray cluster sample} 
\begin{tabular}{| l c c c c c c c c c c |} \hline
name & inst. & exp & $\alpha_{2000}$ & $\delta_{2000}$ & 
$z$ & $\sigma^{\ddagger}$ & ref & 
$T_{\rm X}^{\ddagger}$ & ref & $N_{\rm H}$ \\
 & & sec & hh mm ss & dd mm ss &  & km s$^{-1}$ & opt & keV & X-ray
& $10^{20}$ cm$^{-2}$ \\ \hline
A1631 & IPC & 5705 & 12 52 50 & $-15$ 26 17&  0.0466 & 
628 [{\it 2.7}] & (1) & 2.8 [{\it 653}] & (4) & 3.9 \\
A3528 & PSPC & 15751 & 12 54 18 & $-29$ 01 16& 0.0521 & 
864 [{\it 4.8}] & (2) & ... & ... & 6.1 \\
A3530 & PSPC & 8620 &  12 55 37& $-30$ 21 14& 0.0532 & 
391 [{\it 1.1}] & (2) & 3.2 [{\it 703}] & (4) & 6.0 \\
A1644 & IPC & 11096 & 12 57 15& $-17$ 22 13&  0.0475 & 
933 [{\it 5.6}] & (1) & 4.7 [{\it 869}] & (5) & 4.3 \\
A3532 & PSPC & 8620 & 12 57 19& $-30$ 22 13&  0.0537 & 
594 [{\it 2.4}] & (2) &  4.4 [{\it 838}] & (6) & 6.2 \\
A3556 & PSPC & 13765 & 13 24 06& $-31$ 39 38& 0.0481 & 
554 [{\it 2.1}] & (3)& ... & ... & 4.7 \\
A1736 & IPC & 10660 & 13 26 52& $-27$ 07 33&  0.0446 & 
835 [{\it 4.5}] & (2) & 4.6 [{\it 858}] & (5) & 5.1 \\
A3558 & PSPC & 30213 & 13 27 55& $-31$ 29 32& 0.0475 & 
986 [{\it 6.1}] & (3) & 6.2 [{\it 1012}] & (7) & 4.4 \\
A3559 & PSPC & 8127 & 13 29 54& $-29$ 31 29& 0.0461 & 
279 [{\it 0.6}] & (2) & ... & ... & 4.5 \\
A3560 & PSPC & 14929$^{\dag}$ & 13 31 51& $-33$ 13 25& 0.0462 & 
403 [{\it 1.2}] & (2) & ... & ... & 4.3 \\
A3562 & PSPC & 20202 & 13 33 32& $-31$ 40 23& 0.0491 & 
825 [{\it 4.4}] & (2) & 3.8 [{\it 772}] & (6) & 4.2 \\
A3564 & PSPC & 8011 & 13 34 22& $-35$ 13 21& 0.0498 & ... & (2) 
& ... & ... & 4.5 \\
A3566 & PSPC & 7357 & 13 38 59& $-35$ 33 13& 0.0497 & ... & (2) 
& ... & ... & 4.3 \\
A3571 & PSPC & 6072 & 13 47 29& $-32$ 51 57& 0.0391 & 
1060 [{\it 7.0}] & (2) & 7.6 [{\it 1133}] & (6) & 4.0 \\ 
SC1327-312 & PSPC & 50415$^{\dag}$ & 13 29 45 & $-31$ 
36 12& 0.0495 & 676 [{\it 3.1}] & (3) & ... & ... & 4.3 \\
SC1329-313 & PSPC & 20202 & 13 31 36 & $-31$ 48 45 & 0.0469 &
1044 [{\it 6.8}] & (3) & ... & ... & 4.2 \\ \hline
\multicolumn{11}{l}{(1) Zabludoff \etal 1990; (2) Quintana \etal 1995; 
(3) Bardelli \etal 1994} \\
\multicolumn{11}{l}{(4) White \etal 1996; 
(5) David \etal 1993; (6) Edge \etal 1990;
(7) Day \etal 1991} \\
\multicolumn{11}{l}{ $^{\dag}$ the exposure is equal to the sum of the 
overlapping images} \\
\multicolumn{11}{l}{ $^{\ddagger}$ in the square brackets and italic font,
we report the reference values ($\sigma_{\rm ref}$, $T_{\rm ref}$) obtained
from }\\
\multicolumn{11}{l}{the relationships (cf. White \etal 1996): $T_{\rm
ref}=(0.093 \pm 0.060) \sigma^{(1.831 \pm
0.291)}$ and $\sigma_{\rm ref}=(3.697 \pm 0.547) 
T_{\rm X}^{(0.552 \pm 0.088)}$,} \\
\multicolumn{11}{l}{ where $\sigma$ (measured in 100 km s$^{-1}$) and 
$T_{\rm X}$ (in keV) are the observed values. }\\
\multicolumn{11}{l}{} \\
\multicolumn{11}{l}{{\bf A1631} seq. 1900;} \\
\multicolumn{11}{l}{{\bf A3528} seq. wp300093; also known as Klemola 21
(Klemola 1969), it is a X-ray double cluster:} \\
 & \multicolumn{10}{l}{A3528n ($12^h 51^m 38.2^s$, $-28^\circ 44' 01''$,
$T_X$ = 3.8 keV [{\it 772} km/s] ref.~4); } \\
 & \multicolumn{10}{l}{A3528s ($12^h 51^m 58.5^s$, $-28^\circ 57' 20''$,
$T_X$ = 4.0  keV [{\it 795} km/s] ref.~4). } \\
\multicolumn{11}{l}{{\bf A3530} seq. wp701155n00; its ROSAT field 
contains also A3532.} \\
\multicolumn{11}{l}{{\bf A1644} seq. 7654; it is a cD cluster.} \\
\multicolumn{11}{l}{{\bf A3532} seq. wp701155n00; also Klemola 22.} \\
\multicolumn{11}{l}{{\bf A3556} seq. wp800375n00} \\
\multicolumn{11}{l}{{\bf A1736} seq. 7653; a foreground group of galaxies
is also seen optically at $\sim$ 10400 km/s (Dressler \& Schectman
1988)}\\
\multicolumn{11}{l}{{\bf A3558} seq. wp800076; Shapley8, a cD cluster 
(cf. also PSPC analysis by Bardelli \etal 1995)} \\ 
\multicolumn{11}{l}{{\bf A3559} seq. rp800285n00.} \\
\multicolumn{11}{l}{{\bf A3560} seq. wp800381+wp800381a01+wp800381a02.} \\
\multicolumn{11}{l}{{\bf A3562} seq. rp800237n00.} \\
\multicolumn{11}{l}{{\bf A3564} seq. rp800288n00+a01; no significant
X-ray emission; not considered in the sample analysed.} \\
\multicolumn{11}{l}{{\bf A3566} seq. rp800286; no X-ray significant
emission; not considered in the sample analysed.} \\
\multicolumn{11}{l}{{\bf A3571} seq. rp800287; it is a cD cluster.} \\
\multicolumn{11}{l}{{\bf SC1327-312} seq. wp800076+rp800237n00; 
singled out optically in (3) as SC1329-314 B and in the X--ray band} \\
& \multicolumn{10}{l}{by Breen {\it et al.} (1994).} \\
\multicolumn{11}{l}{{\bf SC1329-313} seq. rp800237n00; 
singled out  optically in (3) as SC1329-314 A and in the X--ray band} \\
& \multicolumn{10}{l}{by Breen {\it et al.} (1994).} \\
\end{tabular}
\end{table*}

\subsection{Deprojection analysis}

We use PSPC images in the 0.4--2 keV band, which maximizes
the signal-to-noise ratio due to reduced background. 
The images are subdivided into four energy bands (42--51, 52--90, 
91--131, 132--201 PI channels) and corrected by the corresponding 
exposure maps using an implementation of
the procedure written by Snowden and collaborators (1994) 
in the Interactive Data Language. The 0.8--3.5 keV
IPC band images (corresponding to the range 5--10 PI channels) are 
also flat-fielded.

The cluster surface brightness profiles are extracted using the XIMAGE 
software (vers.~2.53), and then
analysed using the  deprojection technique pioneered by 
Fabian \etal (1981). As a more detailed description is given by
White \etal (1996), we only summarize the assumptions
and the limitations of this analysis method here.
Assuming spherical geometry for the cluster and hydrostatic 
equilibrium, the deprojection technique enables the volume count
emissivity of the hot ICM to be determined as a
function of radius from the surface brightness profile. 
Including the effects of absorption by the intervening
material (hydrogen column density $N_{\rm H}$) and detector characteristics,
a choice of the gravitational potential (functional form, core 
radius $r_{\rm c}$, velocity dispersion $\sigma_{\rm dpr}$) allows the
physical properties of the ICM to be determined (assuming the perfect gas
law). In this analysis, we use a true isothermal sphere to describe the 
shape of the gravitational potential (Binney \& Tremaine 1987; 
a comparison of the effect of 
different potential laws is described by White \& Fabian 1995).

Given the complexity of the Shapley region,
some uncertainties in the available values of $\sigma$ from optical 
analysis may arise
from diffuse clusters overlapping and/or the ill-defined boundary
of each structure. 
Because of this, we interpolate the $\sigma_{\rm dpr}$ values from 
$T_{\rm X}$ where available, using the general relationships from
White \etal (1996) (see notes of Table~1; sometimes
there can be a significant
discrepancy between the interpolated $\sigma_{\rm dpr}$ and the optical
velocity dispersion, i.e. A3530 and A3532).
The optical velocity dispersion is used
when $T_{\rm X}$ is unknown (i.e.for A3556, A3559, A3560,
SC1327-312 and SC1329-313).
The consistency of the optical value is then compared
against the values inferred by interpolating the deprojected
$T_{\rm X}$ and luminosity $L_{\rm X}$.
If there is a large disagreement, then $\sigma_{\rm dpr}$ is altered
according to the interpolated values (as found for A3559, A3560,
SC1329-313).

In Table~2, the input values
of $r_{\rm c}$, the velocity dispersion $\sigma_{\rm dpr}$ and 
the outer hydrostatic pressure, $P_{\rm out}$, are presented. These
parameters enable us to produce a flat deprojected temperature profile
for each cluster, according to the isothermal assumption. The table also
shows the results of the deprojection analysis. 
The rebin factor gives the degree of grouping of the surface
brightness profile to improve the counts statistics,
and the 
radius $R_{\rm out}$ is the outer extremity of the deprojection 
where the luminosities and masses (the gravitating mass, $M_{\rm dpr}$, 
the gas mass, $M_{\rm gas}$, and the stellar mass, $M_*$) are quoted.
The values with the subscript ``500" are discussed below in Sect.~3.

Note the background contribution is estimated by 
considering the surface brightness value just
outside the maximum radius of each deprojection.

\begin{table*}
\caption{Deprojection analysis results} 
\begin{tabular}{| l c c c c c c c c c c c|} \hline
name & $\sigma_{\rm dpr}$ & $r_{\rm c}$ & $P_{\rm out}/10^4$ & rebin & 
$R_{\rm out}$ & $L_{\rm X,bol}$ & $M_{\rm dpr}$ & $M_{\rm gas}$ 
+ $M_\star$ & $R_{500}$ & $M_{500}$ & $f_{500}$ \\ 
& km s$^{-1}$ & Mpc & K cm$^{-3}$ & $\times$ bin$^{\dag}$ 
& Mpc & $10^{44}$ erg s$^{-1}$ & $10^{14} M_{\odot}$ & $10^{13} 
M_{\odot}$ & Mpc & $10^{14} M_{\odot}$ & $\times100$\%   
\\ \hline
A1631 & 628 & 0.2 & 0.4 & 5 & 0.40 & 0.1 & 0.99 & 0.17 + 0.42 & 
1.23 & 3.07 & 0.060 \\
A3528n & 772 & 0.3 & 4.2 & 3& 0.37 & 1.1 & 1.09 & 0.62 +  $\clubsuit$ & 
1.65 & 7.23 & 0.079 \\
A3528s & 795 & 0.2 & 4.0 & 3& 0.37 & 1.4 & 1.40 & 0.62 +  $\clubsuit$ & 
... & ... & ... \\
A3530 & 703 & 0.4 & 1.0 & 3 & 0.76 & 1.0 & 2.27 & 1.61 + 0.50 &
1.40 & 4.69 & 0.117 \\
A1644 & 856 & 0.5 & 4.0 & 3 & 0.61 & 2.5 & 2.10 & 1.78 + 0.32 &
1.70 & 8.12 & 0.120 \\
A3532 & 791 & 0.4 & 1.5 & 3 & 0.90 & 2.9 & 3.45 & 3.33 + 0.66 & 
1.55 & 6.38 & 0.136 \\
A3556 & 644 & 0.6 & 0.4 & 4 & 1.55 & 0.7 & 4.17 & 3.82 + ... & 
1.29 & 3.63 & 0.102 \\
A1736 & 835 & 0.7 & 3.5 & 3 & 0.63 & 1.4 & 1.62 & 1.45 + 0.27 &
1.66 & 7.54 & 0.129 \\
A3558 & 986 & 0.6 & 2.8 & 3 & 1.03 & 9.8 & 5.57 & 6.84 + 1.20 &
1.94 & 12.18 & 0.176 \\
A3559 & 511 & 0.7 & 0.2 & 3 & 1.19 & 0.3 & 1.97 & 1.89 + ... & 
1.00 & 1.66 & 0.104 \\
A3560 & 738 & 0.5 & 0.7 & 3 & 1.34 & 2.3 & 4.67 & 5.36 + ... &
1.48 & 5.43 & 0.135 \\
A3562 & 772 & 0.6 & 1.0 & 3 & 1.24 & 3.9 & 4.50 & 5.86 + 0.75 &
1.54 & 6.12 & 0.152 \\
A3571 & 1060 & 0.3 & 4.0 & 3 & 0.91 & 14.5 & 6.22 & 5.82 + 0.46 &
1.99 & 12.79 & 0.142 \\ 
SC1327-312 & 676 & 0.5 & 2.2 & 3 & 0.60 & 1.1 & 1.32 & 1.26 + ... &
1.36 & 4.27 & 0.137 \\
SC1329-313 & 511 & 0.4 & 1.0 & 3 & 0.45 & 0.3 & 0.61 & 0.46 + 0.28 &
1.07 & 2.05 & 0.146 \\ \hline 
\multicolumn{12}{l}{$^{\dag}$ the original binsize is $15''$ in PSPC 
images and $16''$ in IPC; $15'' \sim 20$ kpc at $z\sim 0.05$.} \\
\multicolumn{12}{l}{$\clubsuit$ for this double cluster, we estimate that
$M_\star = 0.66 \times 10^{13} M_{\odot}$ at $R$ = 0.74 Mpc, which is the
radius that encloses both }\\
\multicolumn{12}{l}{substructures.}\\ 
\end{tabular}
\end{table*}

\section{THE BARYON FRACTION}

Comparison of observations of light-element abundances 
with standard primordial nucleosynthesis theory indicates that
the baryon density parameter $\Omega_{\rm b}$ lies between 
``reasonable" values of
0.037 and 0.088  $h_{50} ^{-2}$ (see Copi \etal 1995 for a discussion on the
meaning of ``reasonable" due to the fact that the main uncertainties 
are not described by a Gaussian). These limits represent 
$\Omega_{\rm b}^{\rm low}$ and $\Omega_{\rm b}^{\rm high}$ respectively,
that we consider hereafter when referring to the baryon 
density parameter from primordial nucleosynthesis. It should be noted that
these changes, from the previously
accepted value of $0.05\pm0.01$ (Walker \etal 1991), arise from
revised constraints on the D + $^3$He (for the lower 
limit on $\Omega_{\rm b}$) and $^7$Li abundance (for the upper limit).

Thus, if regions that collapse to form rich clusters in an Einstein-de 
Sitter Universe retain the same 
value of $\Omega_{\rm b}$ as the rest of the Universe, 
then only a few per cent of cluster masses can be due to
baryons (mostly gas in the ICM, but also stars in galaxies). 
However, observations of clusters reveal a baryon 
fraction, $f_{\rm b}$ (the sum of the gas fraction $f_{\rm gas}$
in the ICM and the stellar fraction $f_\star$ contributed by the 
cluster galaxies), of 10--30 per cent. 

Historically, X-ray 
observations have always shown a relatively high baryon fraction 
in clusters (e.g. Stewart \etal 1984), although White \& Frenk (1991)
highlighted the discrepancy when new tighter and lower constraints 
from nucleosynthesis were published. In the Coma cluster, 
White \etal (1993; cf. also Briel \etal 1992) found the ratio between the
gas plus stellar
mass to the total gravitational mass to be about 0.226 $h_{50}^{-1.5} +$0.015.
Thus, the disagreement between the observed baryon density and that predicted 
by cosmic nucleosynthesis appears to be exceeded by a factor of $\sim$ 3 for 
all reasonable values of the Hubble constant (there is 
only a weak dependence on $H_0$). 

This problem, sometimes known as the ``Baryon Catastrophe'', 
is not unique to the 
Coma cluster as the discrepancy appears to be similar, if not worse,
in many other clusters (Henry \etal 1993,
White \& Fabian 1995). The baryon content of the Shapley 
Supercluster has also been estimated to be significantly high
at 18 per cent, in a 37 $h_{50}^{-1}$ Mpc region around the core (Fabian
1991). Such an overdensity on this large a 
scale means that the severest problem is then the 
accumulation of baryons within a standard cosmological model.

Possible solutions to the ``Baryon Catastrophe'' are: 
(a) a non-zero Cosmological Constant $\Lambda$, so that $\Omega_{\rm 0,tot} =
\Omega_0 + \Omega_{\Lambda} = 1$ (cf. Carroll \etal 1992 for a discussion
on the limit on $\Lambda$), (b)
a different estimate of $\Omega_{\rm b}$ from the nucleosynthesis calculations
of the light elements abundances (as may appear from the recent 
conflicting observations of deuterium abundances, cf. Hata \etal 1996
and Schramm \& Turner 1996, and references there),
(c) inappropriate X-ray modelling of clusters which may result in 
overestimations of gas mass and/or underestimations of gravitational mass
(e.g. Gunn \& Thomas 1996).

In our case, the gas fraction, $f_{\rm gas}$, is estimated 
by the ratio of gas mass determined in the deprojection analysis
to the gravitational mass. To estimate the stellar fraction, 
$f_\star$, we adopt the optical luminosity function as 
obtained by Raychaudhury \etal (1991) for the 9 clusters and 1 
group in common with our sample. Assuming a mass-to-light ratio appropriate
for elliptical galaxies of 2.46 $\times (L_{\rm B}/10^{10}L_{\odot})^{0.35} h_{50} 
(M/L)_{\odot}$ (White \etal 1993), we estimate the stellar mass 
using the Schechter luminosity function (1976) with a slope of
$\alpha= -1$. 
As the optical luminosity 
function is quoted at 2 Mpc, $M_\star$ is interpolated to a value at
$R_{\rm out}$, assuming a King profile for the mass distribution.

For those 3 clusters (A3556, A3559, A3560) and the SC1327-312 group lacking
the information required to determine their optical luminosity functions,
we adopt the median statistic as a robust estimator  
of the stellar fractions available. The median value of 2.1 per cent
(note, the whole range is between 0.7--4.6 per cent) is used to
determine $M_\star$ in these four objects. 
The results on $f_{\rm gas}$ and $f_\star$, with
the total baryon fraction  $f_{\rm b} = f_{\rm gas}\times 
h_{50}^{-1.5} + f_\star$, appropriate for individual values of $R_{\rm out}$, 
are shown in Fig.~3.

\begin{figure*}
\hbox{%
\psfig{figure=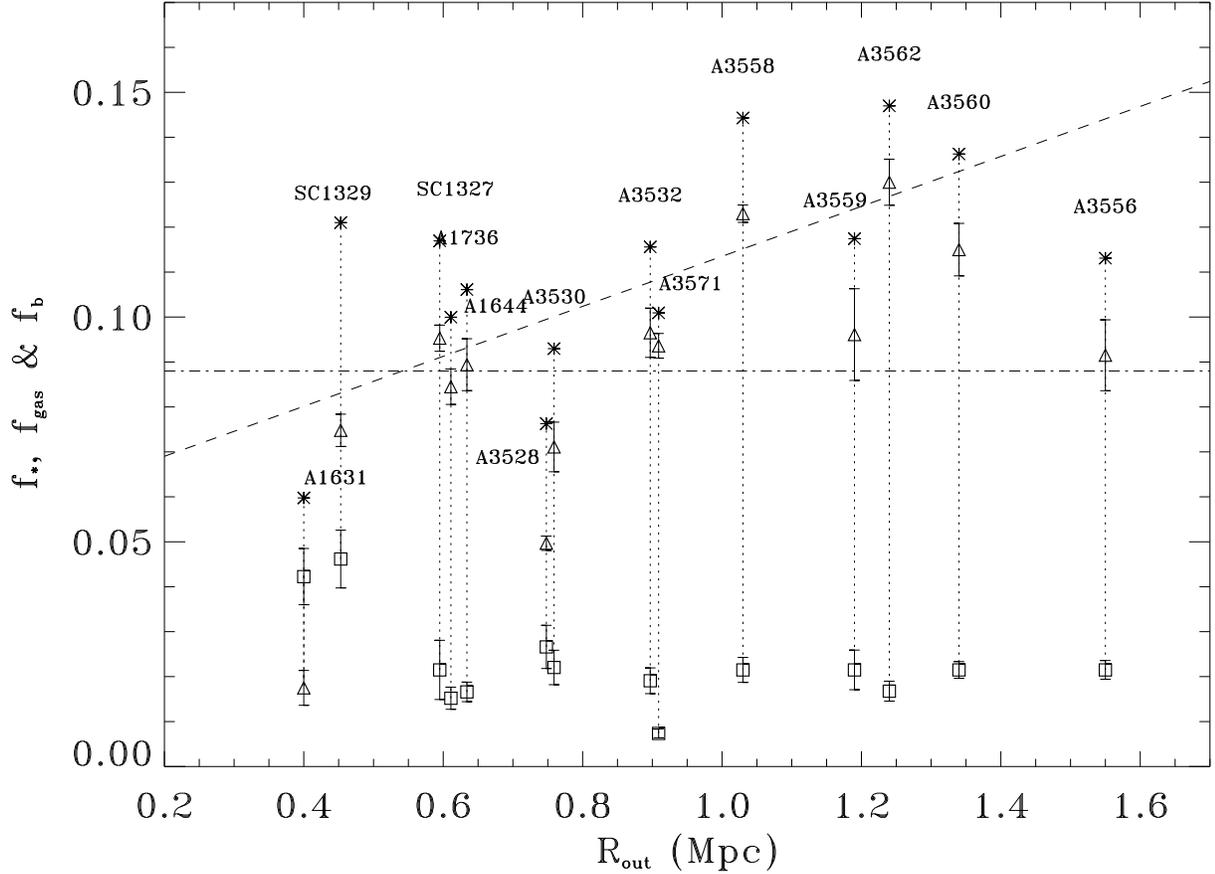,width=\textwidth,angle=90}}
\caption{The baryon fraction $f_{\rm b}$ (asterisk), the gas fraction
$f_{\rm gas}$ (triangle) and the stellar fraction 
$f_\star$ (square), as reported in Table~2, are shown for each cluster in 
our sample. The horizontal dot-dashed line represents the upper limit of
0.088 on $\Omega_{\rm b}$ (Copi \etal 1995), and the
dashed line is the radial $f_{\rm gas}$ dependence reported by 
White \& Fabian (1995): $ f_{\rm gas} = 0.0579 +
0.0556 R$. The errors on $f_{\rm gas}$ come from 
Monte-Carlo replications of the deprojection determinations (on $M_{\rm
gas}$); 
the errors on $f_\star$ are propagated from the luminosity function. 
In these figures the north and south component of A3528 are
considered as one.
The apparent deficit of baryons in A3528 is due to
the different regions used to determine $f_{\rm gas}$ and $f_\star$.
Values for $f_{\rm gas}$ in both of the northern and southern components can 
be determined individually, while $f_\star$ is calculated from the optical
luminosity using a radius which encompasses both components 
(see note $\clubsuit$ in Table~2).
}\end{figure*}

In this figure, a significant
disagreement between the maximum value for $\Omega_{\rm b}$ obtained
from primordial nucleosynthesis analysis (dot-dashed line) and that
determined from X--ray observations of clusters is evident.
This is primarily due to $f_{\rm gas}$, which increases with 
radius, as the contribution of $f_\star$ is almost 
radially independent and consistent with all nucleosynthesis constraints  
(the dispersion around the average value of 0.023 is 0.012).
This emphasizes the problem that any reasonable $f_{\rm gas}$, that is 
$\sim 5 \pm 3$ times $f_\star$ in our sample,
would result in a baryon catastrophe.
 
In order to compare the $f_{\rm b}$ values within a consistent region for 
all clusters, we extrapolate the matter distributions 
to a density contrast equal to 500. The density contrast, $\delta_R$, 
corresponding to a comoving region of radius $R$, is defined as: 

\begin{equation} \delta_R = \frac{(M_{\rm dpr}-\mbox{\bf M}_0)}{\mbox{\bf M}_0},
\end{equation} where $\mbox{\bf M}_0$ is the expected mass in the same
region in a homogeneous universe where the
density $\rho$ is equal to $\rho_c \Omega_0$:
\begin{equation}
\mbox{\bf M}_0 = \frac{H_0^2}{2G}R^3 \Omega_0 \sim 2.9 \times 10^{11}
R^3_{\rm Mpc} \Omega_0 h_{50}^{-1} M_{\odot}. \end{equation}

The overdensity value of 500 is chosen as representative of the cluster
region where the assumption of hydrostatic equilibrium used in
deprojection is still valid. In fact, various numerical simulations
(cf. Evrard \etal 1995, Bartelmann \& Steinmetz 1996) show the gas 
to be largely thermalized inside the relaxed regions characterized
by this overdensity, with outer envelopes that accret and/or merge 
on the cluster at lower density contrast.

Thus, to determine the baryon fraction $f_{500}$, we need 
to extrapolate $M_{\rm dpr}$, $M_{\rm gas}$ and
$M_*$ to the radius $R_{500}$, where $\delta_{R} =500$.
The resulting values of $R_{500}$ have a median value
of 1.54 Mpc and lie between 1.00 (A3559) and 1.99 (A3571) Mpc.
The extrapolated  $M_{\rm dpr}$, $M_{500}$ (cf. 11th column in
Table~2), is 
obtained using an isothermal profile with the
$\sigma_{\rm dpr}$ and $r_c$ as given in Table~2;
the baryonic mass at $R_{500}$ is calculated (i) by fitting a corrected
King model with $\beta$ equal to 2/3 to the
deprojected density profile and integrating the best--fit to obtain the
gas mass, and (ii) extrapolating
the stellar mass as discussed above. 
The last column in Table~2 
quotes the calculated values of $f_{500}$ at the same
density contrast for each cluster (cf. also Fig.~4). 
We obtain a median estimate of
$f_{500} = 0.112 h_{50}^{-1.5} + 0.018 $, for the extrapolated
gas and stellar fraction, respectively, with a range extending from 
0.060 (A1631) to 0.176 (A3558). 

\begin{figure}
 \psfig{figure=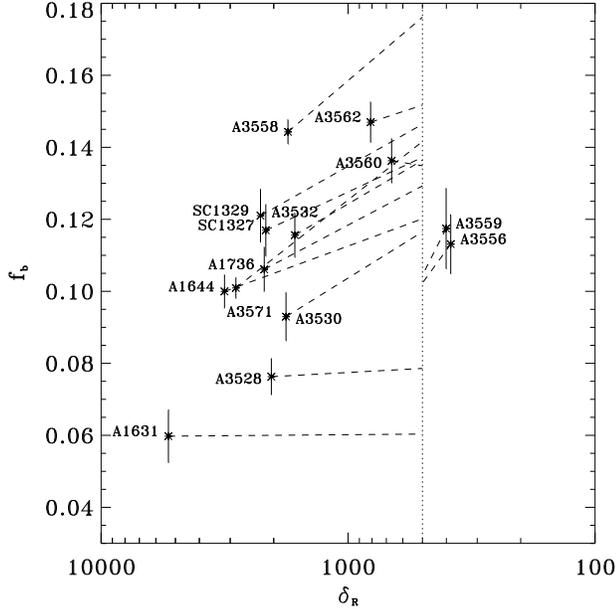,width=0.5\textwidth,angle=0}
\caption{This plot shows the estimated baryon fraction, $f_{\rm b}$, 
and its extrapolated value to $\delta_R = 500$, $f_{500}$,
for the clusters in our sample. The dashed lines connect
the 2 values for each cluster. The vertical
dotted line indicates the density contrast equal to 500.
} \end{figure}

\subsection{The Uncertainties in the Baryon Fraction}

The uncertainties inherent in the single-phase deprojection analysis are 
discussed by White \& Fabian (1995). We note that the method 
gives direct and tight constraints on the gas mass through 
the comparison with the 
observed emission-weighted temperature. Also, any 
multiphase analysis of the ICM in a cooling flow will not 
lower the estimated baryon 
fraction at radii larger than the cluster core (Gunn \& Thomas 1996).
Therefore, the main uncertainties arise from the determination of the 
total gravitational mass. 

Our estimates of the total mass as extrapolated values at density contrast
of 500, $M_{500}$, were obtained using optical velocity dispersions 
(and X-ray temperatures) measured within $R_{500}$ and
adopting the isothermal assumption valid on the so-defined regions.
However, given the complexity of the Shapley Supercluster because of the 
great number of extended sources present in the field,
some uncertainties could arise on the total amount of mass present in each
cluster. To limit this uncertainty, we check
the cluster masses against optical constraints. 
Quintana \etal (1995) have applied the Virial Theorem to 
8 of the 12 clusters and 1 of the 2 groups in our sample (cf. their
Table~7 for estimates on $M_g$ from galaxies within 2 Mpc, which is an
acceptable limit for the thermalized region as 
above shown). Using a linear dependence of mass with radius,
we compare our values of $M_{500}$ with their rescaled $M_g$. 
Generally, the optical value is subject to large errors, both because 
of the unknown shape of the 3--dimensional tensor
of the velocity dispersion and because of projection effects. In fact 
(e.g. see David \etal 1995), there may be discrepancies 
of order a factor 2 between the cluster mass determinations from 
the application of the Virial Theorem to optical data and the results 
from X-ray analysis. Assigning a gravitational radius equal to $R_{500}$, 
we determine a median ratio, with respect to
the X--ray value, of 1.2  (5th and 95th percentile: 0.7 and 2.3).
The extreme cases (A3530, A3559, A3560 at the lower end, and A3556 and
A3562 at the upper end) reflect the disagreement between the 
velocity dispersion obtained optically and that inferred from $T_{\rm X}$.
 
Here we note, and then show in Fig.~5, that only 
by a correction factor larger than 1.5 on the gravitating mass,
can the baryon fraction $f_{500}$ drop down to the upper
limit from the nucleosynthesis calculation (1.48=0.130/0.088, for
$h_{50}=1$; for $h_{50} > 1$ the factor increases till to 3.82 for 
$H_0 = 100$ km s$^{-1}$ Mpc$^{-1}$).

In Fig.~5, we consider corrections that 
could theoretically  reconcile the estimates of $\Omega_{\rm b}$ 
as measured in clusters with the value from the primordial 
nucleosynthesis calculations:
(a) a contribution to the gravitating mass, although it appears
unable to make the necessary correction (when it is less or equal to a
factor 1.5 as discussed above); (b) $\Omega_0 \sim 0.3$, which allows 
agreement within the constraints on $\Omega_{\rm b}$, for all values 
of $H_0$ in the range 50--100 km s$^{-1}$ Mpc$^{-1}$; 
(c) a combination of the 2 above corrections, requiring $H_0>78$ km 
s$^{-1}$ Mpc$^{-1}$ to be consistent with the $\Omega_{\rm b}^{\rm low}$.

However, note that any conclusion on the median values observed in 
our sample become more dramatic when the more luminous clusters such as
A3558 and A3571 are considered. Fig.~4 emphasizes that there is a real 
dispersion in the baryon content of the clusters.

\begin{figure}
\psfig{figure=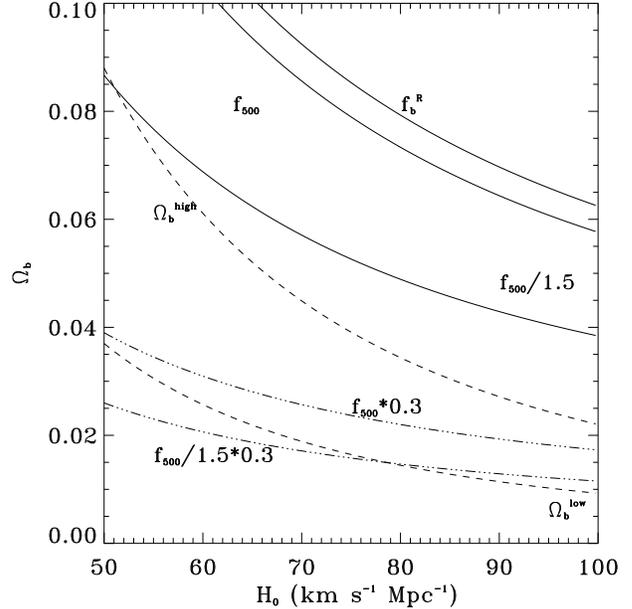,width=0.5\textwidth}
\caption{Constraints on the Hubble constant $H_0$ and the
differently estimated baryon fraction: (dashed line) the upper and lower
limit, $\Omega_{\rm b}^{\rm low}$ and $\Omega_{\rm b}^{\rm high}$, on the 
baryon density parameter from the primordial nucleosynthesis calculations
of Copi \etal 1995; (solid line) the median baryonic contribution on the 
total mass for the clusters in our sample, with and without a correction
factor of 1.5 on the gravitating mass and $\Omega_0=1$; (dot--dot--dash
line) the previous case assuming $\Omega_0=0.3$.
The thick solid line indicates the baryon
fraction $f_{\rm b}^{R}$ observed in the 3 structures defined in Sect.~4. 
} \end{figure}

\section{DISCUSSION}

Using the isothermal gravitational mass profiles, extrapolated to the 
density contrast of 500, $M_{500}$, we attempt
to find how much of the Shapley Supercluster is bound.
Then, the baryonic component of the structures are studied and
the significance of their total mass assessed in terms
of Gaussian random fluctuations in the universal mass density field.  

\begin{figure}
\psfig{figure=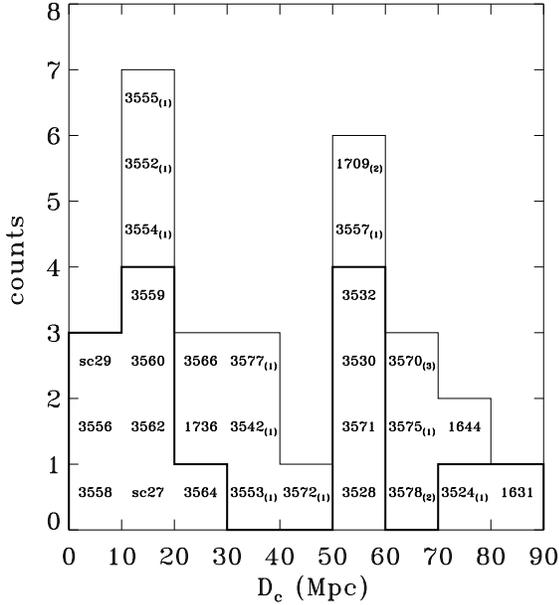,width=0.5\textwidth,angle=0}
\caption{The histogram of the counts of ACO clusters $vs.$ their comoving 
distance from A3558 is shown, overplotted with a thicker histogram of
the counts of the X-ray observed clusters (also in bold font for 
emphasis). Within each bin, the clusters are labelled in increasing
comoving distance, $D_{\rm c}$, upwards.
The redshift measurements are quoted in the present paper and in: (1)
Quintana
\etal 1995; (2) Postman \etal 1992; (3) Abell \etal 1989. 
}\end{figure}

In this analysis on the mass component distribution inferred from 
X-ray observations of clusters, we use the sample defined above and
collect all the data relevant to this aim.
Furthermore, to provide information on the completeness of our sample,
all the clusters present in the Shapley region, as obtained by cross--correlating
the lists from Zucca \etal (1993), Einasto \etal (1994) and Quintana \etal 
(1995), are binned with respect to their comoving separation $R$ from A3558
and shown in Fig.~6. The comoving radius $R$ is defined as
\begin{equation}
R=\sqrt{D_{\rm c,A3558}^2 + D_{\rm c,i}^2 
-2D_{\rm c,A3558}D_{\rm c,i}\cos\theta},
\end{equation}
where $\theta$ is the angular apparent separation on the sky between 
A3558 and the cluster $i$, and $D_{\rm c}$ is the respective comoving 
distance.

As it is 
clear from this figure, not all the clusters in the Shapley region were 
detected or observed in X-ray band. Within 30 Mpc of A3558, 
there are 5 clusters undetected: 3 from PSPC observations (A3554, A3564,
A3566) and 2 (A3552, A3555) from {\it GINGA} scan of the region (Day
\etal 1991). On the other hand, beyond 30 Mpc, 8 out of 10 
clusters were not observed; the other 2, A3572 and A3575, 
were, but remain undetected. 

We also checked for any feature of the X-ray emission from these 15 clusters in 
the {\it ROSAT} All-Sky Survey Bright Source Catalogue (Voges \etal 1996).
Within 0.5$^{\circ}$ of the optical centre as reported in ACO catalogue, 
three (A3524, A3557, A3572) have a suspected hard and extended 
counterpart, an other two (A3564 and A3577) have a hard but not-extended nearby 
source. From the respective  counts rate as quoted in the catalogue, 
and a conversion factor of $1.8 \times 10^{-11}$ erg cm$^{-2}$ s$^{-1}$ 
for 1 count s$^{-1}$, we obtain only for A3572 a 
luminosity of about 3.5 $\times 10^{44}$ erg s$^{-1}$ (corresponding 
to around 2.5 $\times 10^{14}$ M$_{\odot}$ in M$_{\rm gas}$, when its luminosity
is scaled to the Coma cluster), whereas the other 4 clusters have an
estimated luminosity of 0.1--0.3 $\times 10^{44}$ erg s$^{-1}$, equivalent
to M$_{\rm gas} \sim$ 0.6--0.9 $\times 10^{14}$ M$_{\odot}$. 

To summarize, we do not expect a significant contribution from
X-ray detectable baryonic matter extra within radii less than 30 
Mpc from A3558. 
Some conclusions on related topics, such as
possible corrections to our estimates of the baryonic and 
total mass on scales larger than 30 Mpc, will be discussed 
in section 4.1 and 4.2, respectively.

We analyse the distribution of matter that we observe in 4 hierarchical
spherical structures, increasing in size from the core 
(A3558, A3556, A3559, A3560, A3562, SC1327-312 and SC1329-313; Fig.~7)
to the whole supercluster, of comoving radius $R$ centred on A3558 and 
given by eq.~(3).

\begin{figure}
\psfig{figure=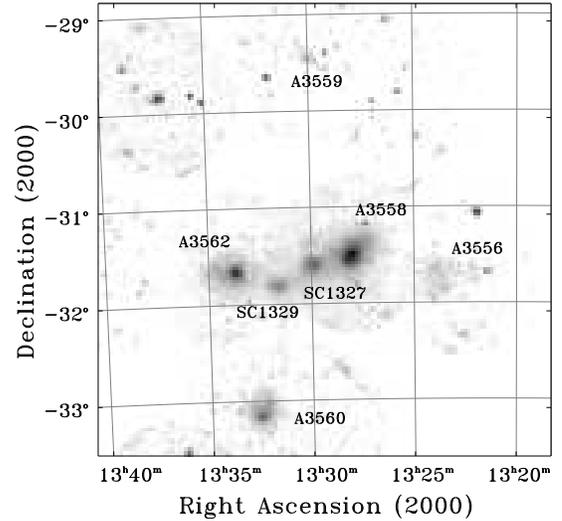,width=0.5\textwidth,angle=0}
\caption{Mosaic of the core of the Shapley Supercluster using all the 
PSPC observations covering the region.
}\end{figure}

We see that the core, structure (1), has a total mass of 
$3.5 \times 10^{15} M_{\odot}$ equal to the sum of
$M_{500}$ for the 5 clusters and 2 groups enclosed within a scale 
of 13 Mpc.  The next structure, (2), is defined as the core and 
A1736, with a total mass of 4.3 $\times 10^{15} M_{\odot}$ within 28 Mpc.
On a scale of 54 Mpc, structure (3), which contains the core, A1736, A3571 
and the western extension (A3528--A3530--A3532), has a total mass is 
$7.4 \times 10^{15} M_{\odot}$.
Finally, the Supercluster as a whole, structure (4), extends to a scale 
of 88 Mpc, with a total mass of $8.5 \times 10^{15} M_{\odot}$. 
Note that these mass estimates are highly conservative limits on the total
mass in each cluster. No account has been made of the mass 
in the outer region of the clusters, where either the isothermal or 
the hydrostatic equilibrium assumption could fail. Extrapolating
$M_{\rm dpr}$ to $\delta=200$ (where the cluster reaches 
virialization), increases the masses by a median factor of
1.49 (in a range 1.40 -- 1.93) with 
respect to $M_{500}$, enlarging by the same factor the mass 
estimates for each of the 4 structures.

A different approach for estimating the expected mass in these structures,
considered as self-gravitating systems, 
is to apply the Virial Theorem (Heisler \etal 1985) on the clusters. As
shown by Raychaudhury \etal (1991), the region is retarded from the cosmic 
expansion through mutual attraction. 
Assuming the extreme condition that each part of 
the Supercluster is virialized,
enables us to calculate the velocity dispersion as a measure of the
gravitational energy; we obtain 395, 482, 1212 and 1121 km s$^{-1}$, 
for the core,
the structures (2), (3) and (4), respectively. The harmonic radius
in each system is also determined, and gives Virial
masses of 19, 35, 367, 400 $\times 10^{14} 
M_{\odot}$, respectively. Again it appears that the core has 
enough gravitating mass to account for the Virial value, as does
structure (2).
For the structures (3) and (4), the Virial mass is about 
5 times larger than the respective $M_{\rm grav}$.

Adopting the value of $\sim 4 \times 10^{16} M_{\odot}$ as an estimate of
the mass in the whole supercluster, we find a contribution of 
almost 30 km s$^{-1}$ to
the component towards the Shapley Supercluster optical
dipole, which is around 520 km s$^{-1}$ (Smoot \etal 1991). 
This is less than 6 per cent of this total, in a flat Universe 
with $h_{50}=1$, and requires
an increase in the cosmological factor $h_{50}/\Omega_0^{0.4}$
(i.e. a low density universe and/or higher Hubble constant) if it is
to be more significant (cf. Quintana \etal 1995).

\begin{table*}
\caption{Shapley structures observed on a different scale $R$. For each
structure, we quote: in column N$_{\rm cl}$, the number of clusters analysed,
the number of clusters without X-ray detection, the number of clusters not 
observed at all, and, in brackets, the total number of clusters in the 
catalogue (cf also Fig.~6); the radius $R$ where we
observe the baryon fraction $f_{\rm b}^R$, the ratio between the total 
baryonic (gas and stellar) matter and the total mass
$M_{\rm grav} = \sum M_{500}$; the mass $M_{\rm PN}$ required to observe a baryon 
fraction equal to the predictions from nucleosynthesis; the virial mass 
$M_{\rm vir}$; the mass $\mbox{\bf M}_0$ expected in a homogeneous universe; 
the positive overdensity $\delta_R$ from
eq.~(1) related to the different mass estimated (details in Table~4).
The errors on $f_{\rm b}^R$ are obtained by propagation of
the uncertainties on the stellar (errors on the luminosity function)
and gas masses (using a 20 per cent uncertainty as upper limit obtained
from the extrapolated values when compared with the observed ones).
$\Omega_0$ is assumed equal to 1.} 
\begin{tabular}{| l c c c c c c c c|} \hline
structure & N$_{\rm cl}$ & $R$ & $f_{\rm b}^{R}$ & $M_{\rm grav}$ & 
$M_{\rm PN}$ & $M_{\rm vir}$ & $\mbox{\bf M}_0$ & $\delta_R$ \\ 
 \# &  & Mpc & $f_{\rm gas}+f_\star$ 
 & $10^{14} M_{\odot}$ & $10^{14} M_{\odot}$ &
$10^{14} M_{\odot}$ & $10^{14} M_{\odot}$ &  \\ 
\hline
(1) & 7+0+0 (7) & 13.4 & $0.128 (\pm 0.013) +0.020 (\pm 0.001)$ &35 & 87  & 19 & 7  & 4.05--11.47 \\
(2) & 8+4+0 (12) & 27.8 & $0.125 (\pm 0.011) +0.020 (\pm 0.001)$ &43& 104 & 35 & 62 & 0.66 \\
(3) & 12+6+3 (21) & 54.2 &$0.118 (\pm 0.008) +0.017 (\pm 0.001)$ &74 & 167 & 367 & 463 & ... \\ 
(4) & 14+7+8 (29) & 88.0 &$0.113 (\pm 0.008) +0.018 (\pm 0.001)$ &85 &186 &400 & 1982 & ... \\ 
\hline
\end{tabular}
\end{table*}

\subsection{On the distribution of the baryonic matter}

When we consider the sum of the stellar, gas and gravitating masses  
contributed by each cluster (cf. Table~3), then for
each of the 4 structures we obtain a baryon fraction $f_{\rm b}^R$
greater than $0.13 h_{50}^{-1.5}$ and, in particular, almost 15 per cent
within 28 Mpc. Although this result
disagrees with the predicted range according to the relaxed 
constraint from the standard big--bang nucleosynthesis model 
adopted here (Fig.~5), it is approaching reasonable agreement with the 
low deterium abundance detection recently reported by Tytler \etal (1996) 
of $\Omega_{\rm b} = 0.096^{+0.024}_{-0.020} h_{50}^{-2}$.   
To improve on the previous estimations of the baryon content for the
Shapley core (cf. Fabian 1991, Makino \& Suto 1993), we place tighter  
limits on the stellar contribution and
avoid the use of any empirical relationship among the X-ray quantities by
modelling each cluster separately. 

So far we have neglected any contribution from an intrasupercluster 
medium (ISCM). If we consider the 1$\sigma$ limit on the diffuse
X--ray surface brightness $I = 2.0 \times 10^{-13}$ erg cm$^{-2}$ s$^{-1}$
ster$^{-1}$ (from the {\it GINGA} Large Area Proportional Counter; Day
\etal 
1991), we obtain a 1$\sigma$ upper limit on the ISCM mass, $M_{\rm ISCM}$,
of 0.5, 1.7, 3.6 and 5.0 $\times 10^{14} M_{\odot}$ for structures (1), 
(2), (3) and (4), respectively (using the thermal bremsstrahlung relation,
$M_{\rm ISCM} \propto T^{-1/4} R_{\rm c}^{5/2} I^{1/2}$, and a King model
density distribution with a core radius $R_{\rm c}$, from structure (1), 
of 13 Mpc and a temperature of 1 keV).
These estimates are between 1 and 6 per cent of the corresponding
total gravitating mass, and between 11 and 52 per cent of the total gas
mass observed in clusters. Given this result, if the detection of 
intrasupercluster gas were confirmed, the baryon catastrophe in
these structures (i.e. on scale enclosed between 10 and 90 $h^{-1}_{50}$ Mpc)
would become more severe, increasing the gas component by  
a factor of 1.1--1.5.
We note that {\it GINGA} cannot place limits on any plausible
intergalactic medium of temperature much below 1 keV.

In Fig.~8, we show the dependence with radius, as measured in the
different structures, of the baryon
parameter, $\Omega_{\rm b}=\rho_{\rm b}/\rho_{\rm c}=f_{\rm b}^R
\Omega_0 [\delta_R(M_{\rm grav}) +1]$. This parameter, on scales where
the overdensity produced by $M_{\rm grav}$ with respect to the background
mass $\mbox{\bf M}$ is still important, retains information on both the
baryon fraction and the overdensity $\delta_R$. Hence, the large
baryonic overdensity in the Shapley core indicates accumulation in a
clearly defined potential well, whereas a deficiency is evident on a 
scale larger than 30 Mpc. 
It is difficult to accept that there has been accretion of baryons from 
the outer part to the supercluster core (the crossing time permits such
motion on scales $\ll$ 10 Mpc).
Thus, part of the observed  relative deficit of baryons outside 30 Mpc 
could be due to the incomplete detection in the X-ray band of the total
gas present. If we consider the median values observed in our
sample of 6$\times 10^{14} M_{\odot}$, for M$_{\rm grav}$, and 0.13, for 
f$_{\rm b}$, and a further contribution of a factor 1.5 on 
the total mass when extrapolated at $\delta = 200$, we are able to increase 
the $\Omega_{\rm b}$ values in Fig.~8 to 1.19, 0.23, 0.05 and 0.02 for 
structure (1), (2), (3) and (4), respectively. 

\begin{figure}\psfig{figure=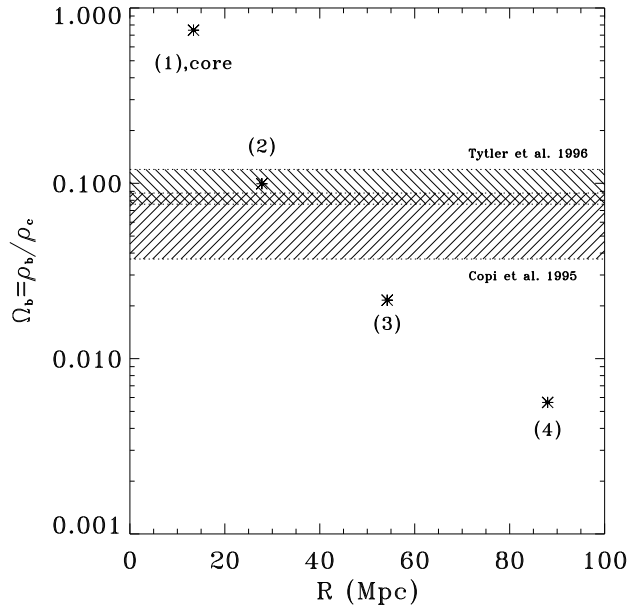,width=0.5\textwidth,angle=0}
\caption{The baryonic parameter $\Omega_{\rm b} = \rho_{\rm b}/\rho_{\rm c}$
is shown as function of the scale of the 4 different structures, which are 
defined in Table~3. The 
two shaded regions indicate  the lower and upper constraints from
the primordial nucleosynthesis as here adopted (Copi \etal 1995) and 
from the recent low deuterium abundance detection (Tytler \etal 1996).
$H_0$ is 50 km s$^{-1}$ Mpc$^{-1}$.
The errors, propagated on the uncertainties of
the baryon density, have dimension comparable to the size of the points.
}\end{figure}

\subsection{On the cosmological consequences}

In Table~3 the mass estimates in a homogeneous universe expected from
eq.~(2) are compared with the total masses extrapolated from the X--ray 
analysis. We note again 
that our mass estimates using $M_{\rm grav}$ are conservative, since we
restricted the determinations to $R_{500}$. A reliable extrapolation to 
$R_{200}$ gives an increase of $\sim$ 50 per cent to the cluster
masses, and thus also to the structures we define. 
Here, we consider another way to constrain the total mass present
in these structures, assuming that all the observed baryons correspond 
to the fraction expected from the primordial nucleosynthesis, i.e.
\begin{equation}
M_{\rm PN}= \frac{M_{\rm b}}{\Omega_{\rm b}} \times \Omega_0 = M_{\rm grav} 
\times \frac{f_{\rm b}^{R}}{\Omega_{\rm b}} \times \Omega_0.
\end{equation} 
In other words, this is the gravitational mass associated with 
a baryonic mass $M_{\rm b}$ for an abundance equal to 
$\Omega_{\rm b}$ in the structures defined above ($M_{\rm b}$ is 
calculated through observations of the gas and stellar matter, and
the value for $\Omega_{\rm b}$ is assumed here to be 0.06, an average
value of the 2 limits from Copi \etal 1995 and within the 2-$\sigma$ range
provided by Tytler \etal 1996). 
In Table~3 we quote $M_{\rm PN}$ for each structure, assuming $\Omega_0=1$.
(Again, an extrapolation to $R_{200}$ increases $M_{\rm PN}$ by a factor 
$\sim 1.5$).

Using $M_{\rm grav}$, $M_{\rm vir}$ and $M_{\rm PN}$ 
as reasonable mass estimates,
we try to assess the probability of obtaining the density fluctuations
related to the structures defined above. The initial distribution of these
fluctuations is assumed to be Gaussian in an Einstein--de Sitter universe. 
Using a spherical collapse model (cf. Appendix A), the initial 
comoving radius $R_0$ of the sphere that contains at the present time 
the mass observed with overdensity $\delta_R$ within 
a comoving radius $R$ is given by: $R_0 = R (1+\delta_R)^{1/3} / (1+z)$. 
The power observed on this scale $R_0$, written with its
dependence on $\Omega_0$ (Peacock \& Dodds 1994, eq. 41), is
\begin{equation}
\sigma_{R_0}^2 = \frac{\Omega_0^{-0.3}}{2\pi^2}\int_{0}^{\infty} 
P_k W^2(kR_0) k^2 dk,
\end{equation}
where $k$ is the wavenumber in units of Mpc$^{-1}$, $W(kR_0)$ is a step
window function that averages out fluctuations on scales
smaller than $R_0$, and the functional form of $P_k$
is the CDM-like power spectrum described by Peacock \& Dodds (1994), 
normalized to the averaged contribution of different classes of galaxy 
and galaxy clusters. It includes corrections for bias, non-linear evolution
and redshift-space distortion effects and shows good agreement with the 
datasets and the COBE results on scales larger than 20 $h^{-1}_{50}$ Mpc.
The same results in the following discussion can be argued 
using a more physical CDM power spectrum,
as approximated in the literature (see e.g. Bardeen \etal 1986) with a 
normalization of $\sigma(16h^{-1}_{50}$Mpc)$\sim 0.6-0.7$, whereas a
forced large-scale normalization to the COBE result gives lower
values for the power by a factor 2, at least on the scales discussed
below (cf. Peacock \& Dodds 1994).
 
To compare the CDM-like power prediction with the overdensities observed 
in the Shapley region, we extrapolate from the non-linear overdensity 
$\delta_R$ to the corresponding present linear value $\delta_{R_0}$
(see for details Appendix A).
Then, the probability that at some point a density fluctuation 
$\delta_0$, with respect to the power on the same scale, exceeds the
observed $\delta_{R_0}$ is given by:
\begin{equation}
P(\delta_0 > \delta_{R_0}) = \frac{1}{\sqrt{2\pi}} \int_{\nu}^{\infty} 
\exp \left( -\frac{t^2}{2} \right) dt,
\end{equation}
where $t=\delta_0/\sigma_{R_0}$ and  $\nu=\delta_{R_0}/\sigma_{R_0}$.

\begin{figure}
\psfig{figure=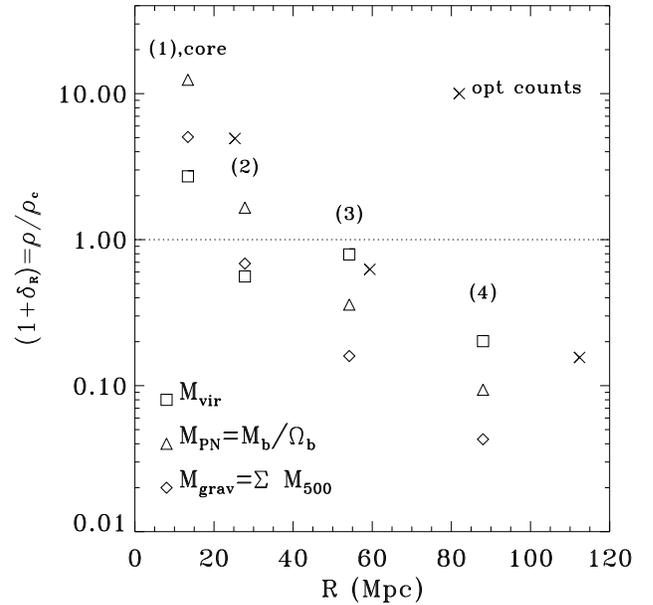,width=0.5\textwidth,angle=0}
\caption{The total mass as sum of the cluster masses, $M_{\rm grav}$, the
expected gravitating mass from the observed baryons, $M_{\rm PN}$, and
the virial mass, $M_{\rm vir}$, are shown as a function of the radius $R$
of the structures which are defined in Table~3 (see text for details). 
The crosses indicate the values extrapolated from the galaxy counts 
(Vettolani \etal~1990, Zucca \etal~1993), assuming that (i) the volumes are
centred on A3558, (ii) a representative mass for the clusters equal to the
median $M_{\rm grav}$ in our sample.  
The dotted line indicates where the density equals the critical density for
$\Omega_0 = 1$. The errors, with dimension similar to the size of the points, 
come from propagating the respective quantities, assuming for the masses
the uncertainties estimated in the deprojection analysis and for 
$\Omega_{\rm b}$ a dispersion of 0.03, consistent with the Copi \etal
result.
}\end{figure}

Using the total mass $M_{\rm grav}$, 
the core is the only structure with a very significant 
overdensity of at least 4.05 and a probability $P$ of $\sim 5$ 
per cent (i.e. 2.0 times $\sigma_{R_0}$, the r.m.s. power value).
Using the overdensity upper limit value of 11.47 (from $M_{\rm PN}$),
structure (1) increases its significance to $P \sim 0.2$ per cent (3.2 
times the corresponding r.m.s. power).
These overdensities suggest that the 
core region is close to the turnaround point, where the perturbed 
region ceases to expand according to the Hubble flow and begins its
collapse. The perturbation has evolved to its maximum expansion 
in $\sim 10-13$ Gyr (the age of the universe in the
present scenario is 13 Gyr)
and its collapsing period will last for almost 10 Gyr more, 
under the cosmological assumptions discussed above.

Structure (2) shows an overdensity of 0.66 within a radius of 28 Mpc 
from A3558, based on $M_{\rm PN}$. 
This corresponds to a probability $P \sim 26$ per cent (1.1 $\times 
\sigma_{R_0}$) and indicates structure (2) as a system 
reaching the turnaround point in about 50 Gyr, before collapsing.

The $\Omega_0$ dependence of $\mbox{\bf M}_0$ and
$M_{\rm PN}$ is linear, whereas 
$M_{\rm grav}$ is independent of it
(i.e. only $M_{\rm PN}$ provides a value of the overdensity
$\delta_{\rm R}$ that does not depend upon any 
assumed cosmological model).
Thus, in a low density universe (e.g. $\Omega_0 \sim$ 0.3) the
overdensities associated with $M_{\rm grav}$ 
increase considerably: in the core and in structure (2) 
$\delta_R$ is equal to 15.8 and 1.3, respectively, 
when measured with $M_{\rm grav}$ (cf. Table~4); structure (3) has 
an overdensity of 1.6, estimating it by $M_{\rm vir}$.
For $\Omega_0 < 1$, the qualitative description discussed above 
on the state of the evolution of the 4 structures is still valid,
given the collapse condition that $\delta > (\Omega^{-1}-1)/(1+z) 
\sim 2.2$ for $\Omega_0=0.3$ (the corresponding condition for 
$\Omega_0=1$ is $\delta > 0$).

Neither structure (3) or (4) exhibits a significant 
density peak, when we use $\Omega_0 =1$.
Similar results can be obtained from the catalogue clusters
counts (Zucca \etal 1993, Vettolani \etal 1990). They describe the Shapley
concentration as the richest supercluster known, with an overdensity, 
with respect to the average density of ACO clusters, of $\sim 10$ 
within a region of 1.4$\times 10^6 h^{-3}_{50}$ Mpc$^3$ enclosing 
25 clusters, to $\sim 400$ for the 9 clusters
contained in the core (as defined by Vettolani \etal 1990). 
This overdensity of counts can be translated in mass overdensities
using the median value for $M_{\rm grav}$ calculated in our
sample, obtaining values in agreement with our estimates as shown
in Fig.~9.

We now return to the issue of the mass of each cluster and of the structures,
which are underestimated. 
Reworking Table~4 using an extrapolation to $R_{200}$, we find
that the core is a 3.7 and 3.5 $\sigma$ fluctuation for $\Omega=1$
and 0.3, respectively, using $M_{\rm PN}$. Only if $\Omega_{\rm b}
\geq 0.095$ does it become less than a 3 $\sigma$ fluctuation.
The 15 undetected clusters (within 90
Mpc) may provide a further contribution of the order of the
observed $M_{\rm grav}$, increasing by a factor 2 the lower limit on the
overdensity (using the median value for $M_{500}$ of about 6 $\times
10^{14} M_{\odot}$). We emphasize that we are assuming the 
structures to have a spherical shape, which is unlikely for large
scale collapsing structures (cf. Peebles 1993; see also 
the conclusion on a ``cigar-like'' shape of the Supercluster by Quintana
\etal 1995 and the recent evidence for a ``bridge'' in the distribution
of galaxies between the core and the western extension; Bardelli, private 
communication). 

\begin{table}
\caption{Significance of the observed positive overdense regions
(i.e. $\delta_R > 0$ for $\Omega=1$, and $\delta_R > 2.2$ for 
$\Omega=0.3$).
The quantities here reported are discussed in the text 
(cf. Appendix A). Starting with observed quantities $R$
and $\delta_R$, we extrapolate the linear value $R_0$ and $\delta_{R_0}$
by eq.~(A4-6) for $\Omega_0=1$ and eq.~(A12-15) for $\Omega_0<1$. The
power $\sigma_{R_0}$ is obtained in eq.(5). 
It can be shown that the correction to $R$ for $\Omega_0 \neq 1$ is not
significant.
Note: the letter $a$ means that the corresponding $\delta_R$ has been  
calculated using $M_{\rm PN}$; $b$, using $M_{\rm grav}$. }
\begin{tabular}{| l c c c c c c c|} \hline
structure & $R$ & $\delta_R$ & $\rightarrow$ & $R_0$ & $\delta_{R_0}$ & 
$\sigma_{R_0}$ & $t_{\rm me}$\\ 
  \# & Mpc &  &  & Mpc &  &  & Gyr\\ \hline 
\multicolumn{3}{c}{$\Omega_0 =1$} &  & \multicolumn{4}{c}{$t_0 = 13.0$ Gyr} 
\\ \hline
(1) $a$ & 13.4 & 11.47 &  & 29.7 & 1.29 & 0.41 & 9.7 \\
(1) $b$ & 13.4 & 4.05 &  & 21.9 & 1.03 & 0.53 & 13.7 \\
(2) $a$ & 27.8 & 0.66 &  & 31.4 & 0.44 & 0.39 & 49.4 \\ 
\hline
\multicolumn{3}{c}{$\Omega_0 =0.3$} & & \multicolumn{4}{c}{$t_0 = 15.8$ Gyr} 
\\ \hline
(1) $b$ & 13.4 & 15.83 &  & 32.8 & 1.33 & 0.45 & 13.9 \\
(1) $a$ & 13.4 & 11.47 &  & 29.7 & 1.28 & 0.49 & 15.9 \\
\hline
\end{tabular}
\end{table}

\section{CONCLUSION}

From our analysis of 12 ACO clusters and 2 groups observed 
with the {\it ROSAT} PSPC and {\it Einstein Observatory} IPC, we have
estimated the baryon 
and total gravitating mass distribution in the Shapley Supercluster.
At a density contrast, with respect to the expected background matter,
of 500 we find that the median baryon fraction observed in the
individual clusters is $f_{500} = 0.11 h_{50}^{-1.5}+0.02$.
This is inconsistent with the standard nucleosynthesis model, 
unless $\Omega_0 \sim 0.3-0.5$ or the low deuterium abundance
is confirmed (from Tytler \etal 1996, the 1-$\sigma$ upper limit on
$\Omega_{\rm b}$ is 0.120). Cluster to cluster, 
there are also a significant difference in $f_{500}$.
In the Shapley Supercluster region, we observe a baryon fraction of 15
per cent within 30 Mpc; a deficiency beyond this
scale is probably due to an undetected warm intergalactic medium.

The gravitating masses $M_{\rm grav}$, calculated as the sum of the
cluster masses within various structures, range from 
3.5 to 8.5 $\times 10^{15} M_{\odot}$, which are lower than the 
minimum necessary to bind the structures 
(i.e. $\delta_R >0$, for $\Omega=1$; 
$\delta_R > 2.2$, for $\Omega=0.3$) except 
the core (A3556, A3558, A3559, A3560, A3562, SC1327-312, SC1329-313).
Applying the Virial theorem to the structure does not 
indicate masses equal the critical one $\mbox{\bf M}_0$.
Then, on scales of 50 Mpc and larger, i.e. for structure
(3) (core--west--A1736--A3571) and (4) (the whole supercluster),
we compute a mass similar to that estimated in a different way by 
Fabian (1991) and Quintana \etal (1995), i.e. $>10^{16} M_{\odot}$. 
Assuming, from primordial nucleosynthesis, a baryon density
parameter of 0.06, we expect a total mass $M_{\rm PN}$
of 0.9 $\times 10^{16} M_{\odot}$ in the core, increasing to 1.8 
$\times 10^{16} M_{\odot}$ for the whole Supercluster. 
These indicate that either structures (3) and (4) are 
actually not bound by a factor of 3 and 10, respectively,
or that we underestimate
the total gravitating mass on scales larger than 30 Mpc (cf. Fig.~9). 

We have used our cluster mass estimates (within $R_{500}$) to assess the 
probability that the various structures within the Shapley Supercluster
arose from gaussian fluctuations in the primordial power spectrum.
Adopting the CDM-like power spectrum of Peacock \& Dodds (1994) and
$\Omega_0=1$, we find that the supercluster core, which has an overdensity
of between 4 and 11, is a fluctuation exceeding 3$\sigma$
which is approaching the turnaround point. If $\Omega_0=0.3$, which
allows the baryon fraction in the clusters to equal that of the rest of
the Universe, then the significance of the supercluster core 
approaches 3$\sigma$ with respect to the power expected on the same scale. 
These significance levels increase to 3.7 and 3.5~$\sigma$ (for $\Omega_0=1$
and 0.3, respectively) if the mass estimates 
are extrapolated to $R_{200}$. 
If the significance of the fluctuations are 
required to be $< 3\sigma$, then $\Omega_{\rm b} \geq 0.095$.
Otherwise, non-gaussian fluctuations need to be considered,
or the power spectrum normalization to be increased on supercluster 
scales. 
    
Further data are needed to make this extrapolation with confidence, but 
the probabilities implied for the structures emphasize the remarkable nature 
of, at least, the core of the Shapley Supercluster. 

\section*{ACKNOWLEDGEMENTS} We are grateful to Harald Ebeling for
his IDL support and to the members of the IoA X-ray Group. 
It is a pleasure to thank our referee, Gianni Zamorani, who has improved 
this work with his suggestions.
ACF acknowledges the support of the Royal Society and DAW that of PPARC.
This research has made use of data obtained through the High Energy 
Astrophysics Science Archive Research Center Online Service, provided 
by the NASA-Goddard Space Flight Centre.

\appendix
\section{A recipe to extrapolate the linear overdensity when $\Omega_0 \leq 1$}
In order to calculate the linear overdensity $\delta_0$ at the present time
and compare it with the observed non-linear overdensity $\delta$ both 
in an open Friedman-Robertson-Walker universe and in an Einstein-de Sitter
universe with cosmological constant equal to 0 and adopting a spherical
collapse model, we refer to sections 11 and 19 of Peebles 
(1980, hereafter Pe), section 8 of Padmanabhan (1993, Pa) and appendix A 
in Lacey \& Cole (1993, LC).

The collapsing region is described in the evolution of its physical
radius $r$ during the time $t$ by the parametric equations
\begin{equation} r = A_c(1-\cos\theta)   \end{equation} 
\begin{equation} t-T_c = B_c(\theta-\sin\theta) \end{equation} 
\begin{equation} A_c^3 = GMB_c^2 \end{equation} 
where $T_c$ will be ignored for a pure growing-mode perturbation (cf. Pa 
eq. 8.21). 

In the case of $\Omega_0=1$, the combination of the 3 equations (A1-3) with
the exact solution of the evolution of the background universe (cf. Pa 
eq. 8.23) and the extrapolated overdensity $\delta_0$ in the linear 
regime (i.e. small $t$) provides the evolution of a spherical overdense region
(cf. Pa 8.32-34):
\begin{equation}
1+\delta = \frac{9(\theta-\sin\theta)^2}{2(1-\cos\theta)^3} \end{equation}
\begin{equation}
t=\frac{1}{2H_0} \left(\frac{5\delta_0}{3}\right)^{-3/2} (\theta-\sin\theta)
\end{equation}
\begin{equation}
r(t) = \frac{3}{10}\frac{R_0}{\delta_0}(1-\cos\theta) = 
\frac{R_0}{(1+\delta)^{1/3}}.
\end{equation}
Of these quantities, the observable at present time $t_0$ are the 
non-linear overdensity $\delta$ and the comoving radius $R=r(t)(1+z)$ allowing
the detemination analytically of $R_0$ and numerically of $\theta$ and 
$\delta_0$. Their values are quoted in Table~4.  

The case for $\Omega_0<1$ is discussed by Pe and a useful 
summary is provided by LC. We present here the corresponding equations
to (A4-6). In a open universe, the background with respect to a collapsing 
region that evolves according to (A1-3) is described by the 
parametric equations (cf. Pe eq. 19.12)
\begin{equation} r_b = A_b(\cosh\eta-1) \end{equation}  
\begin{equation} t = B_b(\sinh\eta-\eta) \end{equation}
\begin{equation} A_b^3 = GMB_b^2.  \end{equation}
Applying the two conditions that we are comparing to a perturbated
overdense
region and a section of the background universe containing the same mass
$M$ and following the same time $t$, we obtain the relations
\begin{equation}
\frac{A_c^3}{A_b^3} = \frac{B_c^2}{B_b^2} \end{equation}
\begin{equation} 
B_c(\theta-\sin\theta)=B_b(\sinh\eta-\eta). \end{equation}
Then, following LC for extrapolation under linear conditions, 
we can write the linear overdensity $\delta_0$ as
\[ \delta_0 = \frac{3}{2} \left[\frac{3\sinh\eta(\sinh\eta-\eta)}
{(\cosh\eta-1)^2}-2 \right] \times \left[1+ \left(\frac{\theta-\sin\theta}
{\sinh\eta-\eta} \right)^{2/3} \right] \]
\begin{equation} \delta_0 = \frac{3}{2} D_t \left[1+ \left(\frac
{\theta-\sin\theta}{\sinh\eta-\eta} \right)^{2/3} \right].
\end{equation}

Thus, a growing density contrast can be described using (A10-11) by
\begin{equation}
1+\delta=\frac{r_b^3}{r^3}=\frac{(\cosh\eta-1)^3}{(\sinh\eta-\eta)^2}
\frac{(\theta-\sin\theta)^2}{(1-\cos\theta)^3}
\end{equation}
\begin{equation}
t=\frac{1}{2H_0}\frac{\Omega_0}{(1-\Omega_0)^{3/2}}\left(\frac{2}{3}
\frac{\delta_0} {D_{t_0}}-1\right)^{-3/2} (\theta-\sin\theta)
\end{equation}
\begin{equation}
r(t) = R_0\left(\frac{2}{3}\frac{\delta_0}{D_{t_0}}-1\right)^{-1}
\frac{1-\cos\theta}{2(\Omega_0^{-1}-1)} = \frac{R_0}{(1+\delta)^{1/3}},
\end{equation}
which represent the equations (A4-6) for $\Omega<1$. 
Furthermore,
it can be shown that the approximation $\Omega_0 \rightarrow 1$, i.e.
$\eta_0 \rightarrow 0$, provided by Taylor expansion of the hyperbolic
functions in (A13), gives the factor 9/2 as required for the case of a
flat universe, i.e. (A4).
To estimate the values quoted in Table~4, we calculate numerically
$\theta_0$ at the present time $t_0$, given $\delta$ and 
$\eta_0 = \cosh^{-1}(2\Omega_0^{-1}-1)$.

Finally,
the time of maximum expansion (i.e. the turnaround point) $t_{\rm me}$ can
be estimated using (A5) and (A14), for $\Omega_0 =1$ and less than 1 
respectively, with $\theta=\pi$.


\begin{thebibliography}{}
\bibitem[]{} Abell, G., Corwin, H.G. \& Olowin, R., 1989, ApJS, 70, 1 (ACO)
\bibitem[]{} Bardeen, J.M., Bond, J.R., Kaiser, N. \& Szalay, A.S.,
1986, ApJ, 304, 15
\bibitem[]{} Bardelli, S., Zucca, E., Vettolani, G., Zamorani, G., 
Scaramella, R., Collins, C. \& MacGillivray, H.T., 1994, MNRAS, 267, 665 
\bibitem[]{} Bardelli, S., Zucca, E., Malizia, A., Zamorani, G., Scaramella,
R. \& Vettolani, G., 1995, A\&A, 305, 435 
\bibitem[]{} Bartelmann, M. \& Steinmetz, M., 1996, MNRAS, 283, 431
\bibitem[]{} Beers, T.C., Geller, M. \& Huchra, J.P., 1982, ApJ, 257, 23
\bibitem[]{} Breen, J., Raychaudhury, S., Forman, W. \& Jones, C., 
1994, ApJ, 424, 59
\bibitem[]{} Binney, J. \& Tremaine, S., 1987, {\it Galactic Dynamics}, 
Princeton University Press
\bibitem[]{} Briel, U.G., Henry, J.P. \& B\"ohringer, H., 1992, AA, 259,
L31
\bibitem[]{} Carroll, S.M., Press, W.H. \& Turner, E.L., 1992, ARAA, 30, 499
\bibitem[]{} Copi, C.J., Schramm, D.N. \& Turner, M.S., 1995, Science, 267, 192
\bibitem[]{} David, L.P., Slyz, A., Jones, C., Forman, W. \& Vrtilek, S.D., 1993, ApJ, 412, 479 
\bibitem[]{} David, L.P., Jones, C. \& Forman, W., 1995, ApJ, 445, 578
\bibitem[]{} Day, C.S.R., Fabian, A.C., Edge, A.C. \& Raychaudhury, S., 1991,
MNRAS, 252, 394
\bibitem[]{} Dressler, A. \& Shectman, S.A., 1988, AJ, 95, 985
\bibitem[]{} Edge, A.C., Stewart, G.C., Fabian, A.C. \& Arnaud, K.A., 
1990, MNRAS, 245, 559
\bibitem[]{} Einasto, M., Einasto, J., Tago, E., Dalton, G.B. \& 
Andernach, H., 1994, MNRAS, 269, 301
\bibitem[]{} Evrard, A.E., Metzler, C.A. \& Navarro, J.F., 1996, 
ApJ, 469, 494
\bibitem[]{} Fabian, A.C., Hu, E.M., Cowie, L.L. \& Grindlay, J., 1981, 
ApJ, 248, 47
\bibitem[]{} Fabian, A.C., 1991, MNRAS, 253, 29p
\bibitem[]{} Gunn, K.F. \& Thomas, P.A., 1996, MNRAS, 281, 1133
\bibitem[]{} Hata, N., Steigman, G., Bludman, S. \& Langacker, P., 1996,
astro-ph/9603087
\bibitem[]{} Henry, J.P., Briel, U.G. \& Nulsen, P.E.J., 1993, A\&A, 271, 413
\bibitem[]{} Kaiser, N. \& Davis, M., 1985, ApJ, 297, 365
\bibitem[]{} Klemola,  A.R., 1969, AJ, 74, 804
\bibitem[]{} Lacey, C. \& Cole, S., 1993, MNRAS, 262, 627
\bibitem[]{} Makino, N. \& Suto, Y., 1993, PASJ, 45, L13
\bibitem[]{} Melnick, J. \& Moles, M., 1987, Rev Mex AA, 14, 72
\bibitem[]{} Mo, H.J. \& White, S.D.M., 1996, MNRAS, 282, 347
\bibitem[]{} Padmanabhan, T., 1993, {\it Structure formation in the 
universe}, Cambridge University Press
\bibitem[]{} Peacock, J.A. \& Dodds, S.J., 1994, MNRAS, 267, 1020
\bibitem[]{} Peebles, P.J.E., 1980, {\it The Large-Scale Structure
of the Universe}, Princeton University Press
\bibitem[]{} Peebles, P.J.E., 1993, {\it Principles of Physical
Cosmology}, Princeton University Press 
\bibitem[]{} Postman, M., Huchra, J.P. \& Geller, M.J., 1992, ApJ, 384, 404
\bibitem[]{} Quintana, H., Ramirez, A., Melnick, J., Raychaudhury, S. 
\& Slezak, E., 1995, AJ, 110, 463
\bibitem[]{} Raychaudhury, S., 1989, Nature, 342, 251
\bibitem[]{} Raychaudhury, S., Fabian, A.C., Edge, A.C., Jones, C. \&
Forman, W., 1991, MNRAS, 248, 101 
\bibitem[]{} Scaramella, R., Baiesi-Pillastrini, G., Chincarini, G., 
Vettolani, G. \& Zamorani, G., 1989, Nature, 338, 562
\bibitem[]{} Schechter, P., 1976, ApJ, 203, 297
\bibitem[]{} Schindler, S., 1996, AA, 305, 756
\bibitem[]{} Schramm, D.N. \& Turner, M.S., 1996, Nature, 381, 193
\bibitem[]{} Shapley, H., 1930, Harvard Obs. Bull., 874, 9
\bibitem[]{} Snowden, S.L., McCammon, D., Burrows, D.N. \& Mendenhall, J.A.,
1994, ApJ, 424, 714
\bibitem[]{} Stark A.A. \etal, 1992, ApJS, 79, 77
\bibitem[]{} Stewart, G.C., Fabian, A.C., Jones, C. \& Forman, W., 1984, ApJ,
285, 1
\bibitem[]{} Tytler, D., Fan, X.M. \& Burles, S., 1996, Nature, 381, 207
\bibitem[]{} Vettolani, G., Chincarini, G., Scaramella, R. \& Zamorani, G., 
1990, AJ, 99, 1709
\bibitem[]{} Voges. W. \etal, 1996, AA, in press
\bibitem[]{} Walker, T.P., Steigman, G., Schramm, D.N., Olive, K.A. \& 
Kang, H.S., 1991, ApJ, 376, 51
\bibitem[]{} White, D.A. \& Fabian, A.C., 1995, MNRAS, 273, 72 
\bibitem[]{} White, D.A., Jones, C. \& Forman, W., 1996, MNRAS, submitted
\bibitem[]{} White, S.D.M. \& Frenk, C.S., 1991, ApJ, 379, 52
\bibitem[]{} White, S.D.M., Navarro, J.F., Evrard, A.E. \& Frenk, C.S., 
1993, Nature, 366, 429 
\bibitem[]{} Zabludoff, A.I., Huchra, J.P. \& Geller, M.J., 1990, ApJS, 74, 1
\bibitem[]{} Zucca, E., Zamorani, G., Scaramella, R. \& Vettolani, G., 
1993, ApJ, 407, 470
\end{thebibliography}
\end{document}